\documentclass[a4paper]{jpconf}
\usepackage{graphicx}

\usepackage{hyperref}
\hypersetup{colorlinks=true,citecolor=blue}
\usepackage{slashed}
\usepackage[]{units}
\usepackage{bbold}
\usepackage{amsmath}
\usepackage{alltt}

\providecommand{\wz}{{\sc Whizard}}
\providecommand{\om}{{\sc O'Mega}}

\providecommand{\tp}{{\sc Toppik}}
\providecommand{\sd}{SINDARIN}

\providecommand{\as}{\alpha_s}
\providecommand{\mts}{M^\text{1S}}
\providecommand{\mtp}{m_t^\text{pole}}

\def\msb{{\overline{\rm MS}}}
\newcommand{\ord}{{\cal O}}

\begin{document}
\title{Top pair threshold production at a linear collider with \wz}

\author{Fabian~Bach, Maximilian~Stahlhofen}

\address{DESY, Notkestra\ss e~85, 22607~Hamburg, Germany}

\ead{fabian.bach@desy.de, maximilian.stahlhofen@desy.de}

\begin{abstract}
We briefly describe how the Monte Carlo generator \wz~2.2 can be employed
to study large QCD effects enhancing the top-antitop production threshold
at a next-generation lepton collider.
While present state-of-the-art predictions at NNLL order are confined to
inclusive total cross sections, our tool can
be used to simulate differential distributions including NLL threshold
resummation in the production, and with off-shell decaying tops.
The new model will be shipped with \wz\ from version~2.2.3 onwards,
to be released along with this article.
\end{abstract}

\section{Introduction}\label{intro}

The scan of the top-antitop threshold at the planned ILC will allow for
a precise determination of the top quark mass, decay width and
couplings. In particular, the expected precision for the top mass will be
an order of magnitude better than what is possible at the
LHC~\cite{Seidel:2013sqa}. The total
cross section for top-antitop threshold production has been calculated
in the framework of NRQCD and is currently known to NNLL
order~\cite{Hoang:2013uda}.
We implement the NLL threshold resummation for the differential
cross section for this process in the Monte Carlo (MC) event generator
\wz~\cite{Kilian:2007gr,Kilian:2014nya,Reuter2014},
allowing for off-shell top quarks and their (tree-level) decay. Our tool
can thus be used for MC studies of exclusive observables or event shapes
in top-antitop production at and above threshold with arbitrary experimental
cuts on the decay products of the tops. Interferences with the
non-resonant tree-level background are automatically taken into account.

\section{\wz\---A universal event generator
for elementary processes at colliders}\label{whizard}

The \wz\ package~\cite{Kilian:2007gr,Kilian:2014nya,Reuter2014}
consists of two components, namely \om~\cite{Moretti:2001zz}
for the generation of the hard parton level matrix elements, and the
\wz\ core program which performs the phase space integration
and delivers all the necessary infrastructure including output formatting
and interfaces. \wz\ comes with a dedicated scripting language, \sd,
to facilitate the user control of the entire package
functionality in one script file, from the process definition in a given model
down to event analysis and histograms/plots.
There is already an extensive library of SM extensions and
BSM models in the package, ranging from various anomalous couplings and
Little Higgs models to SUSY in its
more common flavors as well as exotics like UED etc. Moreover, an interface to
\textsc{FeynRules}~\cite{Alloul2014} exists to further extend this list.
Finally, there are ample package tools and interfaces available to the user
to further control the initial and final state modeling for both hadron and
lepton colliders (LHAPDF, ISR, FSR, MLM~matching, hadronization, beamstrahlung,
etc.) as well as I/O event formats.

Higher-order QCD effects at $t\bar t$~threshold production in
$e^+e^-$~collisions, cf.~section~\ref{theo}, are implemented into \wz\ via
a new model \verb"SM_tt_threshold" described in more detail in
section~\ref{wz_status}.
Along with this proceedings article, a new \wz\ v$2.2.3$ containing the first
official $\beta$-version of the described model will shortly be released at
\begin{alltt}
 \href{https://www.hepforge.org/downloads/whizard}{www.hepforge.org/downloads/whizard}
\end{alltt}

\section{Theory}\label{theo}

\subsection{Top pair production at threshold}\label{thresh}

Near threshold the effective velocity of the top quarks in the center-of-mass
frame, defined as
\begin{equation}
v \equiv \sqrt{\frac{\sqrt{s}-2m + i\Gamma_t}{m}}\,,
\label{veff}
\end{equation}
is of similar size as the coupling, $v\sim \alpha_s\sim 0.1$,
reflecting the bound-state-like dynamics of the top-antitop pair.
To correctly describe top-antitop threshold production it is therefore crucial
to resum Coulomb singular terms $\sim (\alpha_s/v)^n$ in the perturbation series
for the cross section to all orders. 
This Coulomb resummation can be carried out by means of a Schr\"odinger-type
equation within the nonrelativistic effective field theory NRQCD.
The extended vNRQCD framework in addition allows the resummation of large
logarithms of the top quark velocity ($\ln v$).
Schematically the normalized cross section (R-Ratio) close to threshold then
takes the form
\begin{equation}
R = \frac{\sigma_{t \bar t}}{\sigma_{\mu^+\!\mu^-}} = v \sum\limits_k \bigg(\!\frac{\alpha_s}{v}\!\bigg)^{\!\!k} \sum\limits_i (\alpha_s\,\ln\, v )^i \times \bigg\{\! 1 \, ({\rm LL});\;\alpha_s, v \, ({\rm NLL});\;\alpha_s^2,\, \alpha_s v,\, v^2\,({\rm NNLL});\ldots \! \bigg\}.
\label{Rstruc}
\end{equation}
The vNRQCD prediction for the total cross section of top-antitop threshold
production has recently reached the NNLL
level~\cite{Hoang:2011gy,Hoang:2013uda}.
Since $\alpha_{ew} \sim \alpha_s^2$, the electroweak background from
non-resonant production of the top decay products ($W^+b\, W^-{\bar b}$) is suppressed. However, there are effects related
to the interference of double- and single-resonant production that are
parametrically of NLL order~\cite{Hoang:2010gu}. 
For the physical four-particle final state, $W^+b\, W^-{\bar b}$, these
electroweak corrections are automatically
included in cross section predictions produced by \wz, once the (virtual)
top-pair threshold production is consistently implemented.

\subsection{1S mass scheme}\label{mass}

Due to the cancellation of the leading renormalon between the pole mass and the
QCD potential, the threshold production cross section is free of a
$\ord(\Lambda_{\rm QCD})$ renormalon ambiguity, when expressed in terms of a
suitable threshold mass scheme. For the \wz~top-antitop threshold model we use
the 1S mass $\mts$~\cite{Hoang:1999zc} as an input parameter, which can be
related to other short-distance masses, like the $\msb$ mass, in a
renormalon-safe way.
The determination of $\mts$ from a fit to future experimental data will
therefore be stable against higher-order QCD corrections and is not limited
by an intrinsic $\ord(\Lambda_{\rm QCD})$ uncertainty. In fact, a systematic and
statistical error well below \unit[100]{MeV} is expected for the 1S mass
measured by a threshold scan at the ILC~\cite{Seidel:2013sqa}.

\subsection{Top-antitop form factor}\label{FF}

For the implementation of the NLL threshold effects in the \wz~framework we
combine the vector and axial vector production current vertices 
with nonrelativistic S-wave and P-wave form factors, respectively. 
These contain the contributions from the vNRQCD resummation beyond tree level.
The S/P-wave form factors are related to the Green's functions
$G_{1/3}^{(x=0)}(E,p,\nu)$, which depend on the total energy ($E$) of the
top pair as well as on the top three-momentum in the c.m. frame ($p$),
times the vNRQCD Wilson coefficient for the effective spin-triplet S/P-wave
vector current $c_{1/3}(\nu)$~\cite{Pineda:2001et,Hoang:2006ty}. Both, $G_i$ and $c_i$ depend
on the vNRQCD renormalization parameter~$\nu$ and should be evaluated at
$\nu\sim v$ in order to properly resum the logarithms $\sim (\alpha_s \ln v)^n$.
The relevant NLL nonrelativistic momentum-dependent Green's functions can be
computed numerically using the \tp~code~\cite{Hoang:1999zc}.
In the nonrelativistic regime the virtuality of the tops is $\ord(v^2)$.
Hence the NLL nonrelativistic form factor is formally gauge and renormalization
group invariant. This approximation however breaks down for relativistic top
energies and/or momenta and the threshold resummation must therefore be
smoothly turned off in a certain kinematic transition region such that the
cross section matches the correct relativistic continuum obtained from a
corresponding fixed-order QCD calculation.
In a joint effort~\cite{Bach:prep},
we are currently working on a concrete matching procedure between the NLL
threshold and the NLO relativistic continuum, which eventually will be
implemented in \wz.  Until then the new \wz\ model is only reliable in an
energy range of a few GeV centered around the resonance peak at roughly
$2 \mts$.

\section{The \wz\ model}\label{wz_status}

\begin{figure}[t]
\includegraphics[scale=0.59]{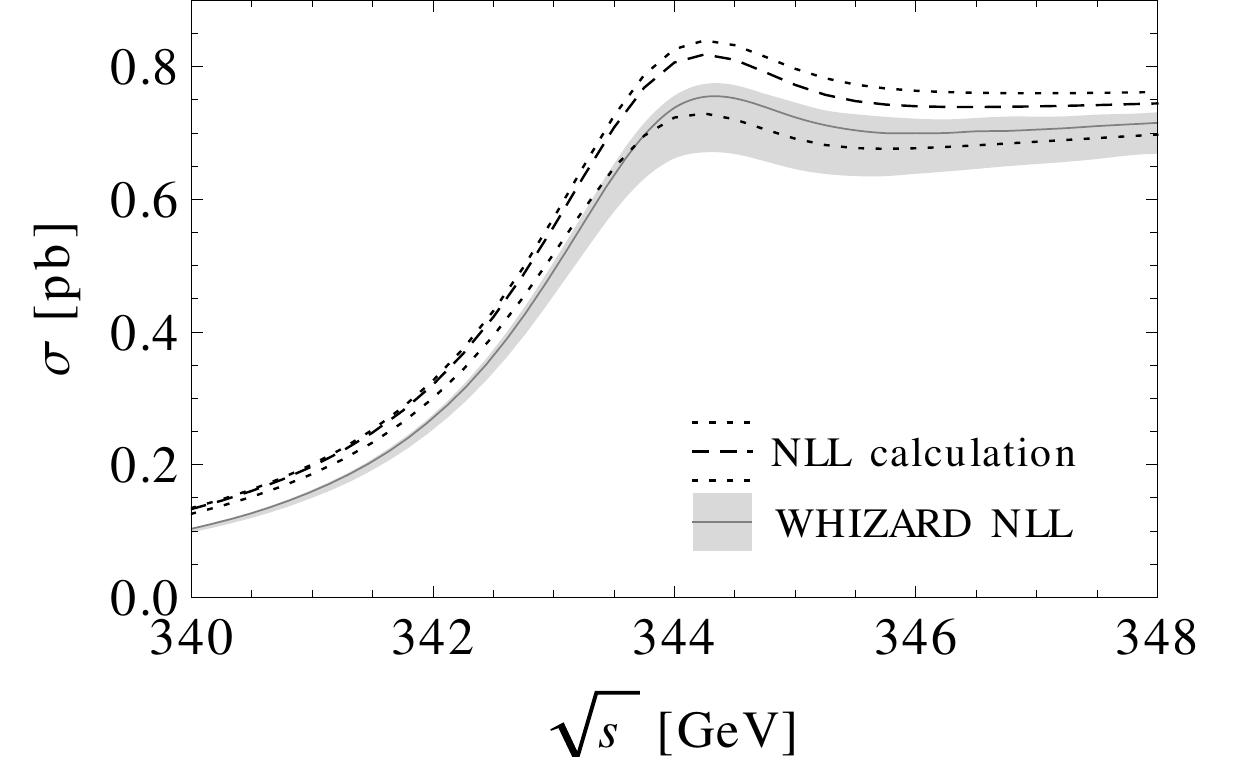}\hspace{3pc}%
\includegraphics[scale=0.59]{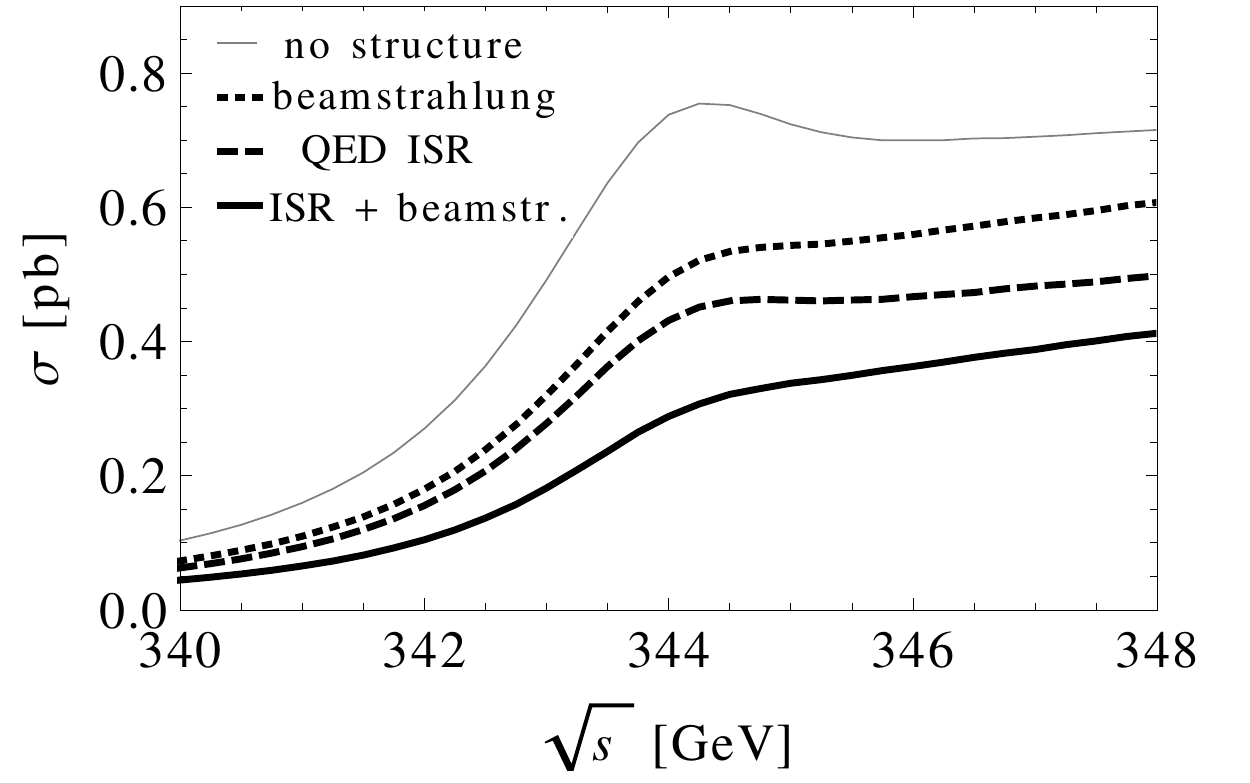}%
\caption{\label{threshold_plots}
Comparison of the vNRQCD and \wz\ inclusive cross sections at NLL,
including error bands from combined scale variations as defined
in~\cite{Hoang:2013uda} (left).
Impact of various beam structure effects on the NLL threshold line-shape
(right). For both plots a top invariant mass cut $\Delta M=\unit[30]{GeV}$
was applied.}
\end{figure}

The nonrelativistic threshold resummation effects are implemented in \wz\
in terms of form factors for the $ttZ/\gamma$ (vector, axial vector) vertices,
as described in section~\ref{FF}.
At each phase space point in the threshold region, the
form factor is calculated as a function of the partonic $\sqrt s$ and the
top 3-momentum magnitude $p$,
either by evaluating an analytical expression at
LL, or by running \tp~\cite{Hoang:1999zc} for NLL precision.
We employ $\mts$ as an independent user parameter for the top mass,
cf.~section~\ref{mass}.
$\mts$ can be converted into the pole mass $\mtp$, which in the current \wz\ version
parametrizes the position of the off-shell top resonances.
Therefore, we compute $\mtp$ internally from the independent model parameters
and provide it as a fixed parameter within \sd, cf.~table~\ref{pars},
in order to facilitate invariant mass cuts on the top resonances.

In the theoretical calculation,
the top width is accounted for by a complex energy, cf.~eq.~\eqref{veff}.
In \wz, however, one can invoke the $2\to 4$ process $e^+e^-\to bbW^+W^-$
including resonant $t\bar t$ diagrams with width $\Gamma_t$ as well as the
full non-resonant background.
Therefore, the four-particle phase space is modeled correctly, and
arbitrary experimental cuts can be applied to the final state.
Although $\Gamma_t$ is implemented as a user-definable parameter,
note that the $2\to 4$
process becomes inconsistent if $\Gamma_t$ is chosen smaller than the
partial decay width $t\to bW^+$ which is fixed by the other model parameters.
\wz\ will warn you whenever this happens.

Finally, the sensitivity of the prediction on the choice for the matching and
renormalization scales can be analyzed by varying the dimensionless
parameters $h$ and $f$, cf.~\cite{Hoang:2013uda}.
In figure~\ref{threshold_plots}, we compare the threshold line-shape at NLL
as produced by the \wz\ model with the results of~\cite{Hoang:2013uda}.
Note the visible difference between the two approaches coming from
relativistic, phase space, and electroweak (interference) effects present
in \wz, but neglected in~\cite{Hoang:2013uda}.
Moreover, as mentioned in section~\ref{whizard}, the user can employ beam
structure like QED~ISR and beamstrahlung in \wz.
The impact on the threshold shape is illustrated in
figure~\ref{threshold_plots}.

\begin{center}
\begin{table}[t]
\caption{\label{pars}
Model-specific \wz\ parameters.}
\centering
\begin{tabular}{@{}*{7}{l}}
\br
Parameter    & Default value     & Description \\
\mr
\verb"alpha_em_i" & $125.924$ & inverse QED coupling $1/\alpha_{em}$ at the $t\bar t$ threshold \\
\verb"m1S" & $\unit[172.0]{GeV}$ & top quark $\mts$ mass \\
\verb"wtop" & $\unit[1.54]{GeV}$ & top quark width $\Gamma_t$ \\
\verb"nloop" & $1$ & NRQCD order: 0 (LL) or 1 (NLL) \\
\verb"sh" & $1.0$ & matching scale parameter $h$ as defined in~\cite{Hoang:2013uda} \\
\verb"sf" & $1.0$ & renormalization scale parameter $f$ as defined in~\cite{Hoang:2013uda} \\
\verb"mtpole" & fixed & top quark pole mass $\mtp$ \\
\br
\end{tabular}
\end{table}
\end{center}

\vspace{-2ex}

\section{Conclusions and outlook}

A precise assessment of the top-antitop threshold cross section will be one of
the key physics cases of a future lepton collider, because it allows for a
precise measurement of the top couplings including $\as$ and, in particular,
the top mass (in a renormalon-free scheme).
While NRQCD theory predictions have recently reached NNLL order
accuracy for the total cross section, we present an implementation of the
threshold enhancement at NLL order for off-shell $t\bar t$ production into the
Monte Carlo generator \wz\ (from version 2.2.3).
We thus provide a fully differential
tool facilitating experimental studies of threshold scans with
arbitrary cuts on the final states of the (LO) decaying top quarks.
In a future version of the model, the threshold
process will be matched onto relativistic continuum production at NLO.


\ack{The authors would like to thank Andr\'e Hoang and J\"urgen Reuter
for many useful discussions.
F.~B.~thanks the TOP2014 organizers for all their efforts and the choice of
the wonderful venue.}


\section*{References}
\providecommand{\href}[2]{#2}
\providecommand{\eprint}[1]{\href{http://arxiv.org/abs/#1}{{#1}}}
\bibliography{references}

\providecommand{\newblock}{}
\begin{thebibliography}{10}
\expandafter\ifx\csname url\endcsname\relax
  \def\url#1{{\tt #1}}\fi
\expandafter\ifx\csname urlprefix\endcsname\relax\def\urlprefix{URL }\fi
\providecommand{\eprint}[2][]{\url{#2}}

\bibitem{Seidel:2013sqa}
Seidel K, Simon F, Tesar M and Poss S 2013 {\em Eur.Phys.J.\/} {\bf C73} 2530
  (\textit{Preprint} \eprint{1303.3758})

\bibitem{Hoang:2013uda}
Hoang A~H and Stahlhofen M 2014 {\em JHEP\/} {\bf 1405} 121 (\textit{Preprint}
  \eprint{1309.6323})

\bibitem{Kilian:2007gr}
Kilian W, Ohl T and Reuter J 2011 {\em Eur.Phys.J.\/} {\bf C71} 1742
  (\textit{Preprint} \eprint{0708.4233})

\bibitem{Kilian:2014nya}
Kilian W, Bach F, Ohl T and Reuter J 2014  (\textit{Preprint}
  \eprint{1403.7433})

\bibitem{Reuter2014}
Reuter J, Bach F, Nejad B~C, Kilian W, Ohl T {\em et~al.\/} 2014
  (\textit{Preprint} \eprint{1410.4505})

\bibitem{Moretti:2001zz}
Moretti M, Ohl T and Reuter J 2001  (\textit{Preprint} \eprint{hep-ph/0102195})

\bibitem{Alloul2014}
Alloul A, Christensen N~D, Degrande C, Duhr C and Fuks B 2014 {\em
  Comput.Phys.Commun.\/} {\bf 185} 2250--2300 (\textit{Preprint}
  \eprint{1310.1921})

\bibitem{Hoang:2011gy}
Hoang A~H and Stahlhofen M 2011 {\em JHEP\/} {\bf 1106} 088 (\textit{Preprint}
  \eprint{1102.0269})

\bibitem{Hoang:2010gu}
Hoang A~H, Reisser C~J and Ruiz-Femenia P 2010 {\em Phys.Rev.\/} {\bf D82}
  014005 (\textit{Preprint} \eprint{1002.3223})

\bibitem{Hoang:1999zc}
Hoang A and Teubner T 1999 {\em Phys.Rev.\/} {\bf D60} 114027
  (\textit{Preprint} \eprint{hep-ph/9904468})

\bibitem{Pineda:2001et}
Pineda A 2002 {\em Phys. Rev.\/} {\bf D66} 054022 (\textit{Preprint}
  \eprint{hep-ph/0110216})

\bibitem{Hoang:2006ty}
Hoang A~H and Ruiz-Femenia P 2006 {\em Phys.Rev.\/} {\bf D74} 114016
  (\textit{Preprint} \eprint{hep-ph/0609151})

\bibitem{Bach:prep}
Bach F, Hoang A, Reuter J, Stahlhofen M and Teubner T {\em in preparation\/}

\end{thebibliography}

\end{document}